\documentclass[journal]{IEEEtran}
\IEEEoverridecommandlockouts
\usepackage{cite}
\usepackage{amsmath,amssymb,amsfonts,amsthm}
\usepackage{algorithmic}
\usepackage{graphicx}
\usepackage{textcomp}
\usepackage{comment}
\usepackage{url}
\usepackage{authblk}
\usepackage{array}
\usepackage{multirow}
\usepackage{subfigure}
\usepackage[ruled,linesnumbered]{algorithm2e}

\usepackage[table]{xcolor}
\usepackage{booktabs}

\usepackage{enumitem} 
\usepackage[table]{xcolor}
\newcommand{\tabincell}[2]{\begin{tabular}{@{}#1@{}}#2\end{tabular}}

\begin{document}

\title{Empowering Computing and Networks Convergence System with Distributed Cooperative Routing}

\author{Yujiao Hu, \textit{Member, IEEE}, Qingmin Jia, Meng Shen, Renchao Xie, \textit{Senior Member, IEEE}, Tao Huang, \textit{Senior Member, IEEE}, F.Richard Yu, \textit{Fellow, IEEE}

\thanks{Yujiao Hu and Qingmin Jia are with Future Network Research Center, Purple Mountain Laboratories, Nanjing 211111, China (email: huyujiao@pmlabs.com.cn, jiaqingmin@pmlabs.com.cn).
Meng Shen is with Pervasive Communication Research Center, Purple Mountain Laboratories, Nanjing 211111, China (email: shenmeng.nj@gmail.com). 
Renchao Xie and Tao Huang are with the State Key Laboratory of networking and Switching Technology, Beijing University of Posts and Telecommunications, Beijing 100876, China (email:renchao\_xie@bupt.edu.cn, htao@bupt.edu.cn). 
F. Richard Yu is with the Department of Systems and Computer Engineering, Carleton University, Ottawa, Canada (e-mail: richard.yu@carleton.ca).

This work was supported in part by the National Nature Science Foundation of China under Grants 92267301 and 92367104. 
}
}

\markboth{Journal of \LaTeX\ Class Files,~Vol.~14, No.~8, August~2021}%
{Shell \MakeLowercase{\textit{et al.}}: A Sample Article Using IEEEtran.cls for IEEE Journals}


\maketitle

\begin{abstract}
The emergence of intelligent applications and recent advances in the fields of computing and networks are driving the development of computing and networks convergence (CNC) system. However, existing researches failed to achieve comprehensive scheduling optimization of computing and network resources. This shortfall results in some requirements of computing requests unable to be guaranteed in an end-to-end service pattern, negatively impacting the development of CNC systems. In this article, we propose a distributed cooperative routing framework for the CNC system to ensure the deadline requirements and minimize the computation cost of requests. The framework includes trading plane, management plane, control plane and forwarding plane. The cross-plane cooperative end-to-end routing schemes consider both computation efficiency of heterogeneous servers and the network congestion degrees while making routing plan, thereby determining where to execute requests and corresponding routing paths. Simulations results substantiates the performance of our routing schemes in scheduling computing requests in the CNC system.
\end{abstract}

\begin{IEEEkeywords}
    Computing and networks convergence; computing requests; future intelligence application; cooperative routing
\end{IEEEkeywords}

\section{Introduction}
Computing and networks have become two pillars of modern social life. 
The networks provide sharing ways to transmit massive data generated by significantly increasing network subscribers and smart devices. 
The computing is responsible for dealing with the massive data according to the expectations of users. 
Currently, the emergence of a large number of intelligent applications and businesses, such as autonomous driving, internet of things, smart manufacturing, augmented/virtual reality, etc., makes the demands for powerful computing and networks (C\&N) capacities more urgent. 

Recall the technology developments of the two pillars. In the early stage, the two pillars kept developing independently and have been upgraded for many generations. 
However, after entering the digital age, the upgrading progresses of the two pillars are slowing down, because the proliferation of users and smart devices makes it difficult for even the most powerful computing devices and the highest bandwidth to meet their computing and data transmission demands.
Until recent years, some technologies that integrate computing and networks gradually break through this dilemma. 
One achievement is software defined network (SDN) \cite{kreutz2014software}. 
Compared with traditional network architecture that forwards data based on routing tables, 
SDN introduces a central management module in the control plane to realize various computing tasks related to dynamic network configuration, and control traffics and resource utilization in the network through combining with specific algorithms. 
Another breakthrough is cloud computing \cite{chen2011cloud}, which provides users with nearly unlimited computing capacity through network accessing. Moreover, cloud computing has been the booster of digital transformation of social life and industries \cite{mohindra2015industry,ke2020massive}.
These influential technological breakthroughs are driving the profound fusion of computing and networks technologies. The government and standards organizations are also encouraging exploration in this direction \cite{9108989}.  

Some researchers have started to explore effective methods to empower computing and networks convergence. 
An integrated framework for software defined networks, caching and computing is proposed \cite{inproceedingschen2017,articlechenToN2018}. The framework defines the processing flows for computing/caching/networking interests respectively, but less considers the operation pipelines for interest packages that require more than two kinds of resources. 
The computing first network \cite{cfn2019icn} is presented to provide compute graphs in an information-centric network (ICN) \cite{xylomenos2013survey} to perform computing load management and performance optimization. However, it focuses on the remote method invocation in ICN and ignores the negative impact of network congestion on computing performance.
Sky computing \cite{yang2023skypilot}, which aims to integrate dispersed cloud resources over networks so that jobs can easily migrate within this vast and heterogeneous collection of commodity clouds, is seen as a future cloud business model. Its subsequent researches, skypilot \cite{yang2023skypilot} and skyplane \cite{jain2023skyplane}, propose novel multi-cloud cooperation computing and high-level data transfer mechanisms, but still focus more on the computation efficiency of different clouds, and consider less about the network transmission. The researches \cite{wang2020net,tang2023siats} suffer from similar issues. 

Motivated by the challenge, this article proposes a distributed cooperative routing framework to empower the convergence of computing and networks. Our novel framework overcomes the barriers of separation between computational routing and network routing, and realizes an end-to-end computing request routing with consideration of both computation efficiency of computing nodes and the degree of network congestion. The end-to-end computing request routing is also capable to ensure the deadline requirements of requests, meanwhile, minimize the money cost of users. Simulations are conducted on the networks with heavy/light network loads. The results show that compared to computing first routing, which tries to guarantee the deadline from the perspective computation level and ignore the effects of network status, our routing framework can better meet the deadline requirements and has lower costs. The main contributions of this article are highlighted as follows:
\begin{itemize}
	\item We illustrate the importance of developing computing and networks convergence (CNC) through analyzing the various forces from social and commercial development, the inevitable trend of technological progress, and the requirements of practical applications. 
	\item We study the connotation of CNC from the perspectives of definition and convergence levels, and proposes a novel routing framework that can support end-to-end deadline-aware computing request routing to empower the development of CNC. 
	\item We deeply analyze the seamless end-to-end request routing scheme to provide distributed cooperation, deadline awareness and cost-friendly completion. 
	\item We conduct simulations on networks with heavy loads and light loads to illustrate the capacity of proposed distributed cooperative end-to-end request routing framework to ensure deadline and save cost.  
\end{itemize}

We organize the remainder of the article as follows. The next section presents the motivations of CNC by analyzing the various forces from social, economic and technical development. Then the connotation of CNC is studied. The proposed computing request routing framework is then described. Following that, the end-to-end cooperation routing schemes based on the framework is analyzed in detail. Afterwards, we show the simulation results to prove the claimed achievements. Finally, we conclude the article.

\section{Why converge computation and networks ? } \label{section:motivations}

\subsection{Social and commercial development requires CNC }
Cisco forecasts by 2023,  global mobile subscribers will grow to 5.7 billion (71 percent of population), global mobile devices will grow to 13.1 billion, and the global fixed broadband speeds will reach 110.4 Mbps \cite{ciscoprediction2020}. The growth will lead to a dramatic increase in data volume, challenging the network transmission bandwidth and in-network storage/computing capacity.
Meanwhile, a large number of intelligent applications have emerged in recent years, including smart transportation, smart manufacturing, Internet of Things (IoT), holographic call, digital twins, AI training and inference, etc. The applications produce massive diverse tasks, such as latency-sensitive tasks, computation-intensive tasks, IO-intensive tasks and normal tasks. All tasks need to be completed on time, indicating huge computing, caching and transmission resources are required. 

At present, the most advanced networking technology (eg 5G) and computing paradigm (eg cloud/edge computing) can deal with business demands in communication or computing, but cannot meet the integrated C\&N requirements. It severely restricts the future development of society and economy. Therefore, a novel C\&N paradigm that is able to ensure on-time response to businesses with customized integrated requirements in the transmission and processing is urgent.

\subsection{The inevitable trend of technological progress}
\textit{Networking. } The traditional networking architecture adopting distributed management is becoming difficult to quickly adapt to changes in network environment, because resource volume is significantly increasing and status of resources are usually time-varying. 
To overcome the problems of poor dynamic adaptability and weak customization ability of distributed architecture, cloudification technology is introduced, that is, a central processing module is deployed in the control plane to process data analysis related to network management and control the network traffics, network functions, routing paths, etc. At present, 5G core network adopts this method and realizes flexible deployment of network functions in data centers at different levels according to business requirements. Deterministic networking (DetNet) ensures a bounded latency and jitter for real-time applications by this way. The future networking management will also benefit from integrating some technologies in the computing field to support more flexible dynamic network configuration.

\textit{Computing. } Since the establishment of data center, the evolution of computing paradigm has begun to rely on the networking. Users have to access corresponding servers/clouds through networking firstly, and then enjoy the computing/storage resources. Moreover, the remote computing heavily depends on the data transmission capability of networking. To provide on-time services, next generation of computing paradigm has to ensure not only sufficient computation resources but also ultra-reliable low-latency data transmission.  

\begin{table*}
	\caption{Descriptions of scenarios and use cases.}
	\renewcommand{\arraystretch}{1.2}
	\begin{center}
		\setlength{\tabcolsep}{4pt}
		    \begin{tabular}{l|l|l|l}
			\hline
			\textbf{Scenarios}  & \textbf{Use cases}  & \textbf{Case challenges}  & \textbf{CNC efforts}  \\ \hline
			\tabincell{l}{Smart \\ Manufacturing \\(SM)} & 
				\tabincell{l}{Multi-AGV cooperative workshop \\ logistics \\ Machine vision based quality \\ control \\ Employee pose recognition \\ Industrial big data mining \\ Remote production control}
				& \tabincell{l}{Many AGV/camera/sensor/human \\ participants \\ Many delay-sensitive tasks \\ Massive data mining and storage }
				& 
    \tabincell{l}{Computing nodes and communication links authorized by \\ SM factories will access to CNC as CNC nodes. Then \\ CNC deploys services required by the production and \\ management processes on CNC nodes. CNC allows multiple \\ tasks to share C\&N resources. According to the requirements\\ of tasks, CNC can quickly locate corresponding CNC \\ nodes and issue the using permissions to specific tasks.}   \\ \hline 
			\tabincell{l}{Vehicle to \\ everything \\ (V2E)}  & \tabincell{l}{ Roadmap updating \\ Intersection passing under vehicle \\ road coordination \\  Vehicle formation  \\ Path navigation  \\  Raw data exchange and \\ high-precision localization } 
				&\tabincell{l}{Large-scale data collection and \\ analysis \\ Latency-sensitive data exchange \\ between vehicles and road \\ Joint C\&N control \\  Remote service supports}  
				& \tabincell{l}{Empowered by CNC, roadside devices and cloud/edge \\ servers are able to provide computing capacity to all V2E \\ tasks. CNC will schedule C\&N resources to transfer and \\ analyze traffic data, ensure steady connections between \\ vehicles and roadside devices.}  \\ 
		\hline
			\tabincell{l}{Augment/virtual \\ reality \\ (AR/VR)}  & \tabincell{l}{AR/VR sightseeing   \\ VR interactive games \\ Image rendering } 
			& \tabincell{l}{High bandwidth
 \\
			Many image processing tasks} 
			& \tabincell{l}{CNC will automatically allocate high bandwidth for AR/VR \\ use cases and schedule enough computing capacity to image \\ processing tasks. }  \\ \hline
		\end{tabular}
		\label{table:descriptions-use-cases}
	\end{center}
\end{table*}

\subsection{CNC will enable a number of use cases } 
CNC is expected to promote many new applications. We presents three scenarios and 15 use cases as representatives in Table~\ref{table:descriptions-use-cases}  to demonstrate the potential of CNC in new applications. 

\section{Connotation of CNC}
\subsection{What is CNC ?}
\textbf{CNC} refers to an approach to heterogeneous resources management that enables
dynamic, flexible and efficient C\&N configuration in order to meet the personalized customized requirements of vast majority of users, improve resource utilization and reduce energy consumption. The main breakthrough of CNC is to transform the independent development of computing paradigm and networking evolution into cooperative improvement. CNC considers not only computing capacity requirements but also the constraints of communication resources while scheduling resources for customized businesses. 

Compared with edge computing that physically brings resources close to network edge to ensure improved response time for latency-sensitive businesses, CNC will logically realize computing resource proximity, that is, through efficient and dynamic network configuration, users feel that the required computing resources can always be satisfied and the requests can be quickly responded to, as if the computing resources were around.
Compared with the network cloudification technology that introduces some high-performance computing equipment/platforms to ensure network management, CNC expands the business scope to C\&N management. 

\subsection{Convergence Levels of CNC} \label{section:convergence-levels}
\begin{figure}[t] 
	\centering
	\includegraphics[width=0.99\columnwidth]{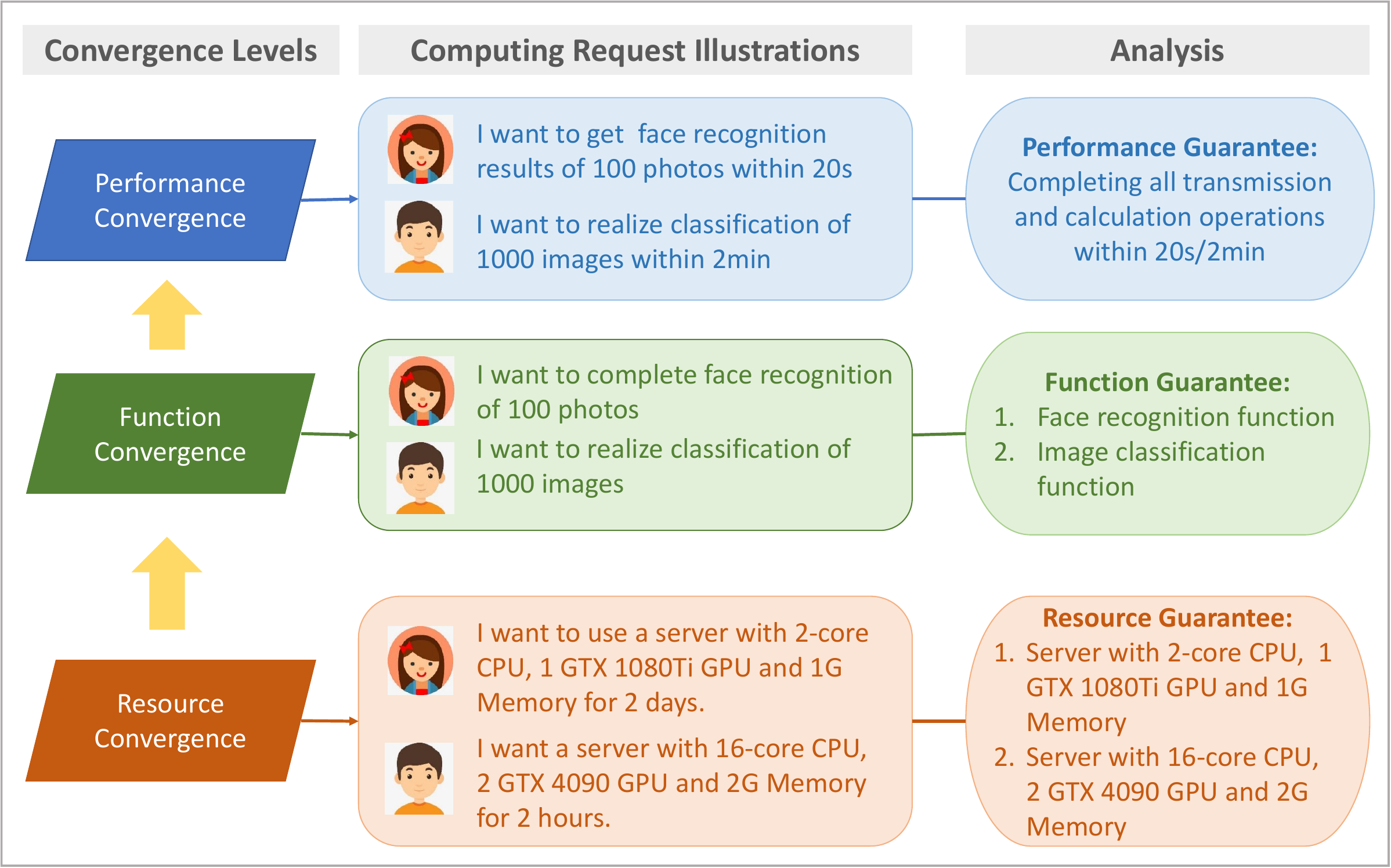} 
	\caption{An illustration for convergence levels of CNC systems: resource convergence, function convergence and performance convergence. }
	\label{fig:CNC-convergence}
\end{figure}

CNC should be able to realize convergence of resource, function and performance. An illustration of three convergence levels is presented in Fig.~\ref{fig:CNC-convergence}. 
\begin{itemize}
    \item Resource convergence focuses more on how to schedule C\&N resources to meet the requirements of tasks in computing capacity. The scheduled C\&N resources must be continuous, i.e. the provided computing resource and the requester must be connected by the scheduled network channel. 
    \item Function convergence pays more attention to allocate clients continuous C\&N resources where the functions required by clients are deployed. The function includes services, platforms, tools, network functions, and so on.
    \item The performance convergence requires that the allocation of continuous C\&N resources ensures the performance requirements of clients, such as completing the request within $20s$ or $2min$. 
\end{itemize}

\begin{figure}[t] 
	\centering
	\includegraphics[width=0.98\columnwidth]{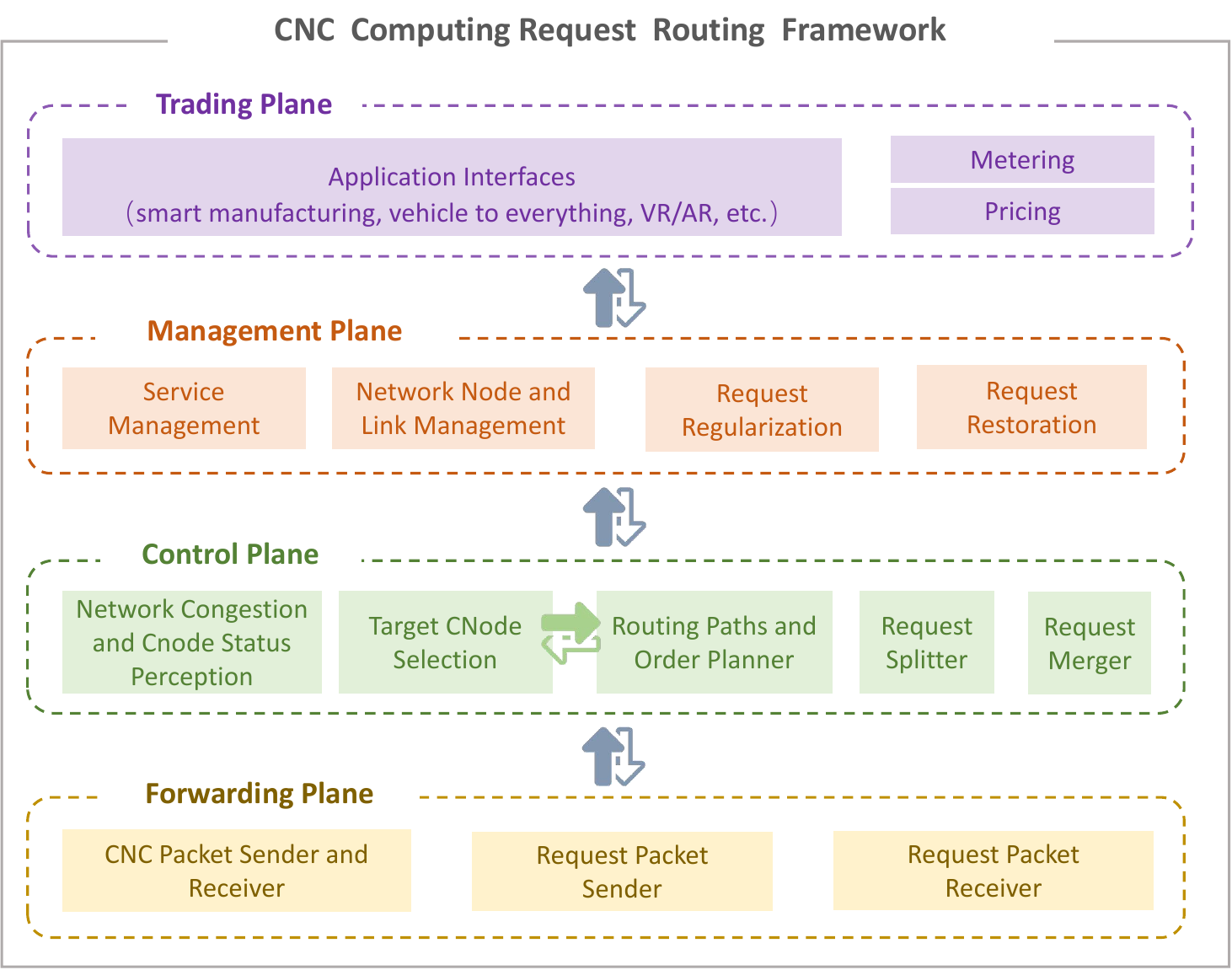} 
	\caption{End-to-end computing request routing framework. }
	\label{fig:CNC-hierarchicalFramework}
\end{figure}

\section{Computing Request Routing Framework}
We propose the computing request routing framework to empower computing and networks convergence.
As presented in Fig.~\ref{fig:CNC-hierarchicalFramework}, the framework consists of four planes: trading plane, management plane, control plane and forwarding(data) plane. The functions of each plane will be explained as follows. 

\subsection{Trading Plane}  
The trading plane provides various application interfaces and corresponding pricing strategies to clients. Specially, it consists of three modules: \textit{Application Interfaces}, \textit{Metering} and \textit{Pricing}. 
The module, \textit{Application Interfaces}, opens various application interfaces to customers, such as intelligent manufacturing, vehicle to everything, AR/VR, etc.
The module, \textit{Metering}, supports the measurement of the usage of computing resources used to complete the computing requests.
The module, \textit{Pricing}, generates corresponding bills for successfully executed computing requests. 

\subsection{Management Plane}
The management plane has three functions. 
Firstly, it manages the status of services deployed in the networks and the congestion degrees of network, in order to provide a real-time global view to support planning in the control plane. 
Secondly, it needs to regularize requests of different convergence levels to a unified representation, to support uniform planning for all computing requests in the control plane. 
Thirdly, it need to revert from the unified expression to the original form of requests, and then provide feedback to clients on the results. 

Four modules are designed to realize the functions in the management planes. 
\textit{Service Management} maintains the real-time information of each service, including the distribution of computation nodes where the service is deployed, and how many replicas are deployed on each computation node and how much backlog that each computation node has. 
\textit{Network Node and Link Management} maintains the real-time status of networks, including the dynamic network topology and the congestion degrees on network routers/switches.
\textit{Request Regularization} is responsible to regularize requests to a unified representation, which is a combination of functional requirements and deadline requirements. Specifically, for a request at the resource convergence level, its functional requirement is to have idle computing resources with specified CPU, GPU, and memory capacity, and its deadline requirement is the specified usage duration. For a request at the function convergence level, its functional requirement is the required service, and its deadline requirement is set to a large value by default. For a request at the performance convergence, its functional requirement is automatically identified based on the request descriptions, and its deadline requirement is specified by clients. 
\textit{Request Restoration} restores the original request from the regularized representation, and return the corresponding execution results to clients. 

\subsection{Control Plane}
The control plane decides the update frequency of service status and network status, where the computing requests are routed and the routing paths.  We designed five modules to support the capabilities of the control plane. 

The module, \textit{Network Congestion and Cnode Status Perception}, determines the frequency of state information exchange with other routers and computation nodes (Cnodes). 
The module \textit {Target Cnode Selection} collaborates with the module \textit {Routing Paths and Order Planner} to determine how to break down requests into multiple sets of tasks, where each set of tasks will be executed, and the routing path for each set of tasks. 
The module \textit{Request Splitter} is responsible to split the request as the decision of \textit {Target Cnode Selection}, while the module \textit{Request Merger} integrates the results of each task set.

\subsection{Forwarding Plane}
The forwarding Plane is responsible for the actual movement of data from one router to another according to the instructions from the control plane. 
The data packets that are forwarded can be divided into two types: CNC packets and computing request packets. 
CNC packets are used to delivery status information of computation nodes and network congestion. 
Computing request packers refer to the data packets of computing requests, such as Internet of Vehicles, cloud games, data anaylis, etc. 
Forwarding and receiving CNC packets are supported by the module \textit{CNC Packet Sender and Receiver}. 
The modules \textit{Request Packet Sender} and \textit{Request Packet Receiver} supports the forwarding and receiving of computing requests, respectively. 

\begin{figure*}[t] 
	\centering
	\includegraphics[width=1.96\columnwidth]{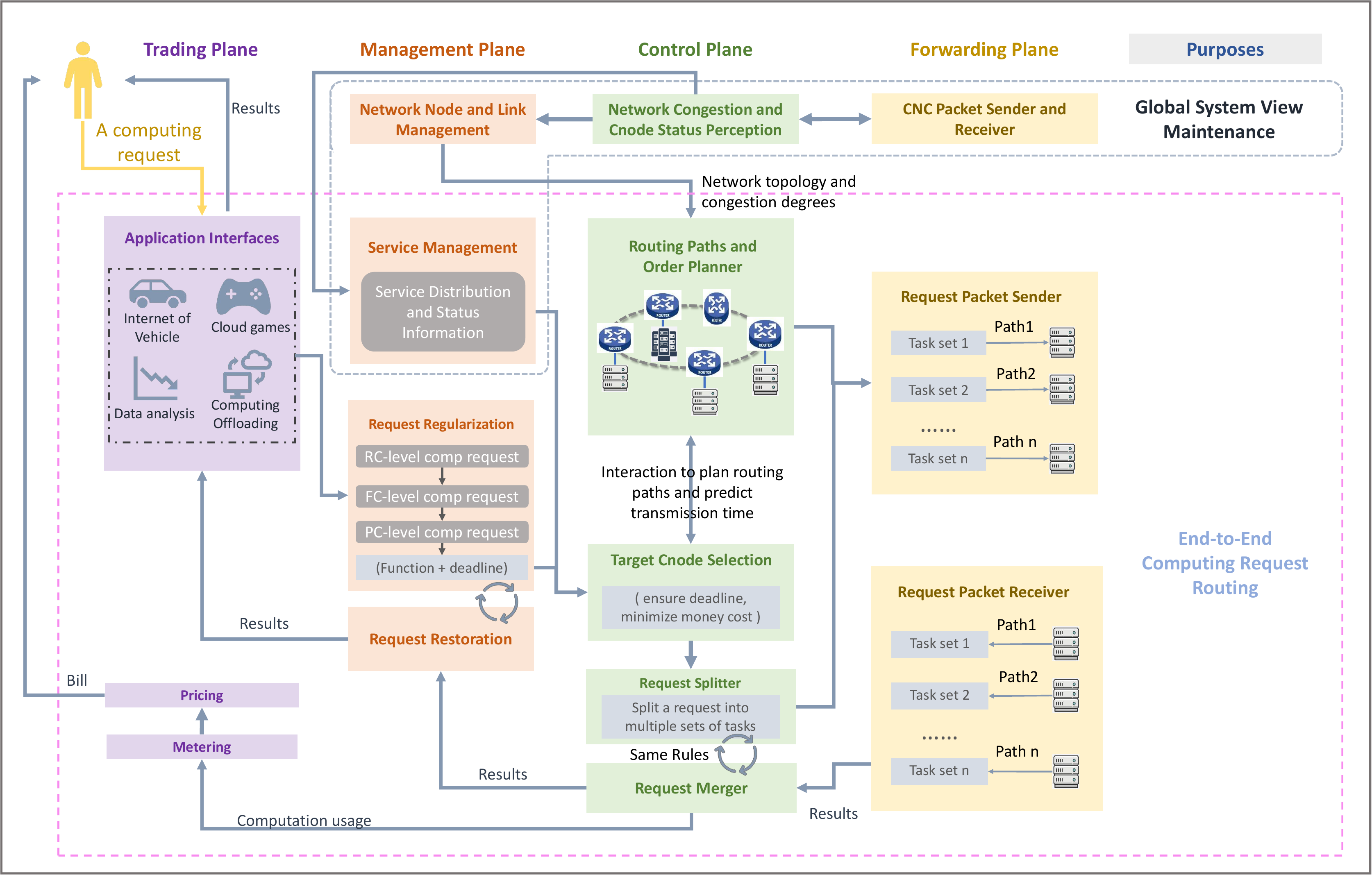} 
	\caption{End-to-end cross-plane cooperative routing schemes based on the computing request routing framework. }
	\label{fig:cross-plane-cooperation}
\end{figure*}

\section{End-to-End Cross-Plane Cooperation Routing}
Based on the computing request routing framework, we can achieve seamless end-to-end computing request routing through cross-plane cooperation. The cooperative routing schemes, as presented in Fig.~\ref{fig:cross-plane-cooperation}, include global system view maintenance and end-to-end computing request routing, which can be divided into request scheduling and result feedback. 

\subsection{Global System View Maintenance}
The maintenance of global system view depends on the cross-plane cooperation among modules \textit{Network Node and Link Management}, \textit{Service Management}, \textit{Network Congestion and Cnode Status Perception} and \textit{CNC Packet Sender and Receiver}. The module \textit{Network Congestion and Cnode Status Perception} decides when to broadcast its computation and network resource status information, and the module \textit{CNC Packet Sender and Receiver} carries out those instructions by forwarding CNC packets. At the same time, the module \textit{CNC Packet Sender and Receiver} also receive CNC packets from other routers and provides these new information to the module \textit{Network Node and Link Management} and \textit{Service Management} through its control plane, in order to update the global system view. 

An accurate real-time global system view can support routers to make optimized routing decisions by considering the entire network topology, traffic patterns, and potential bottlenecks. This can lead to improved performance, reduced latency, and better resource utilization. In our framework, global system view maintenance builds a solid foundation for  end-to-end computing request routing.

\subsection{Computing Request Scheduling}
Clients can submit a computing request to the CNC system through \textit{Application Interfaces}. Then the module \textit{Request Regularization} represents the request in a combination structure of functions and deadline, and transmit it to the module \textit{Target Cnode Selection}. Afterwards, \textit{Target Cnode Selection} obtains global system view from \textit{Service Management} and \textit{Network Node and Link Managements}, and cooperates with \textit{Routing Paths and Order Planner} to get a feasible scheduling decision that can ensure the function and deadline requirements of the request and help save computation cost. This process is the most important part in the end-to-end routing. When planning a feasible routing decision, the module \textit{Target Cnode Selection} must evaluate the execution time based on the current service state and predict the transmission time of each schedule based on the routing path planner, so that the entire response time, including the transmission time in the network and the execution time on the computing node, will not exceed the deadline. Then a cost-friendly schedule is selected from all feasible solutions. With the routing decision, the request is decomposed into multiple task sets in the module \textit{Request Splitter}. Following that, the module \textit{Request Packet Sender} in the forwarding plane forwards each task set to its execution computing node, according to the routing path planned by \textit{Routing Paths and Order Planner}.

\subsection{Computing Result Feedback}
After the scheduled computing node completes the task set, it will return the result, and the module \textit{Request Packet Receiver} is responsible for receiving data packets that encapsulate the result. Then the received results are transmitted to \textit{Request Merger} to merge them together in the original splitting order. Afterwards, the merged results are transferred to the module \textit{Request Restoration} to match the initial form of the request. Finally, the results are fed back to the client through \textit{Application Interfaces}. At the same, the module \textit{Metering} measures computation resource usage and the \textit{Pricing} generates a corresponding bill to the client.  

\section{Simulation Results}
\begin{figure}
    \centering
    \includegraphics[width=0.99\linewidth]{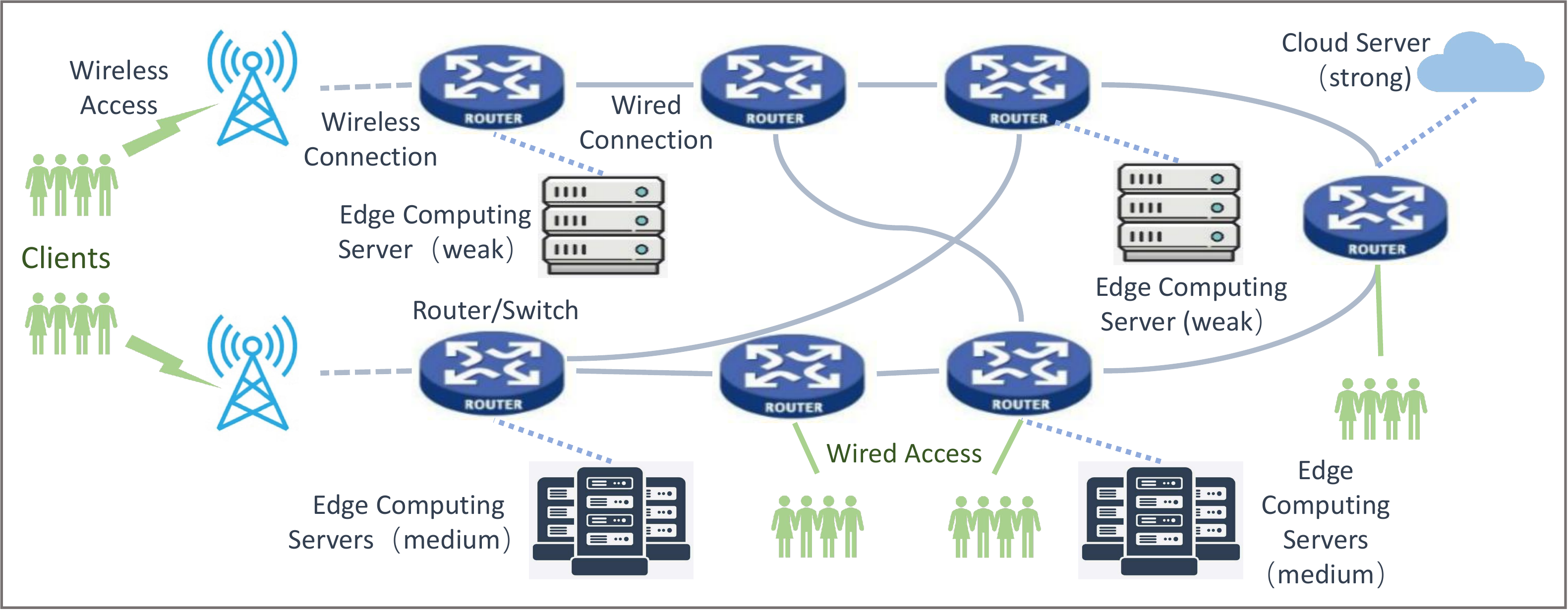}
    \caption{A typical computing and networks convergence system.}
    \label{fig:CNC-system-simulations}
\end{figure}

In this section, we evaluate the capacity of distributed cooperative routing schemes to empower computation and networks convergence through conducting simulations on computing requests with deadlines routing.  In particular, we consider a CNC system structured as Fig.~\ref{fig:CNC-system-simulations}, where the computing servers can interconnect with each other through networks and different servers have varying computational efficiency (weak, medium, strong). Clients can submit computing requests to the CNC system through wireless and wired access. The network bandwidth of the CNC system is set to 1Gbps. The computing requests are generated under the guidance of image processing \cite{pmlr-v162-wortsman22a}. By comparing the success ratio and costs of routing of computing requests in scenarios with heavy/light network loads, experiments can demonstrate the performance of the end-to-end cooperation routing. 

\begin{figure}
    \centering
    \includegraphics[width=0.97\linewidth]{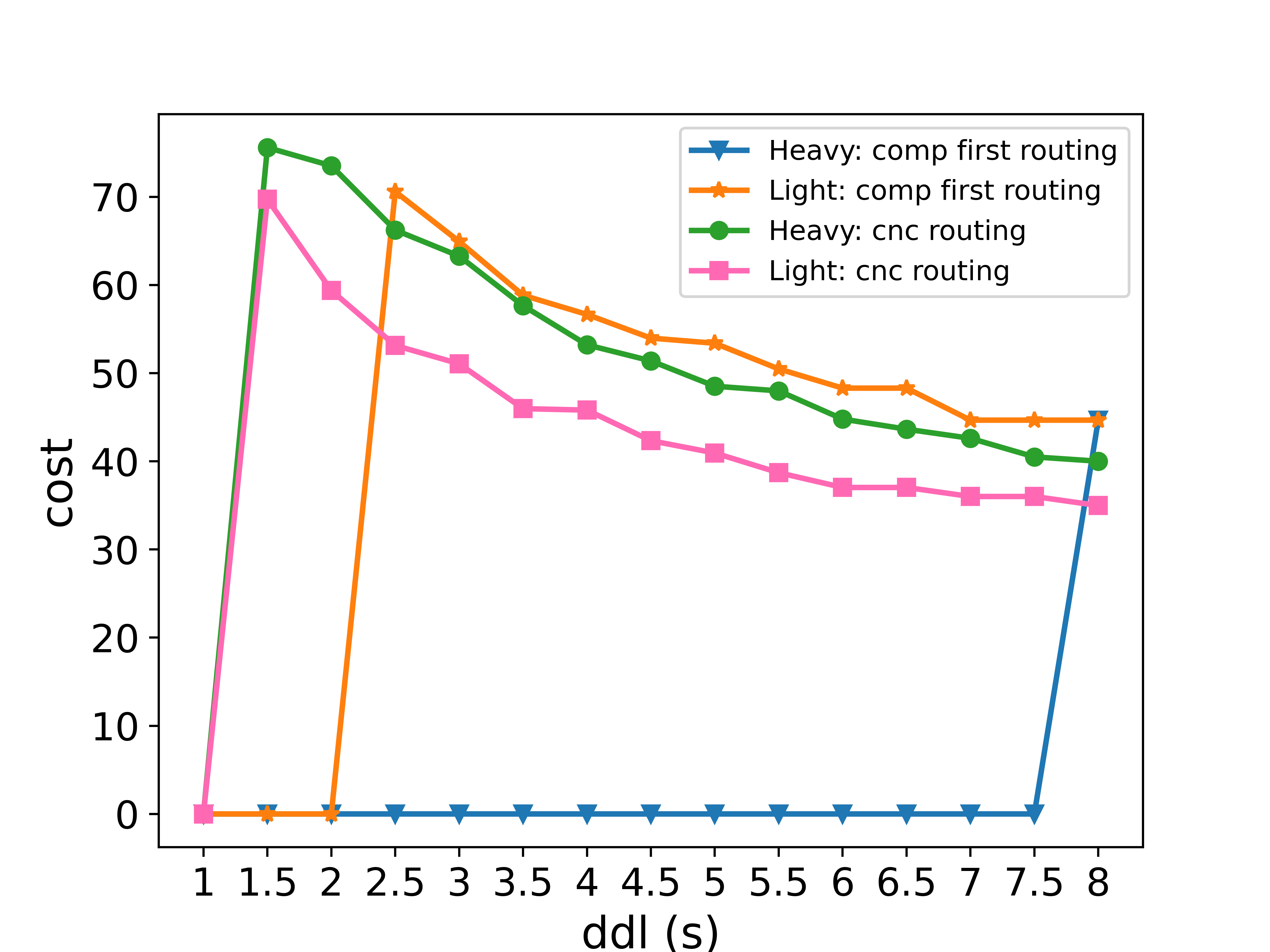}
    \caption{The completion cost of computing requests under different levels of network congestion. The lower the better. But when the cost equals to zero, it means the request cannot be routed under the deadline requirement.}
    \label{fig:simulation-results}
\end{figure}

As shown in Fig.~\ref{fig:simulation-results}, the curves in pink illustrate that the CNC routing with simultaneous consideration of both computation efficiency of heterogeneous servers and network congestion has higher possibility to satisfy the strict deadline requirements and help save costs. Moreover, the cost decreases as the deadline become loose. The heavy network loads have negative effects on the successful ratio of out routing schemes. Meanwhile, a significant negative effect of heavy network loads is experienced with the computing first routing, which consider more about the computation efficiency and roughly estimate the transmission time. Even when the networks are with light loads, the computation cost of computing first routing is higher than the CNC routing. 

\section{Conclusion}
In this article, we propose end-to-end computing request routing schemes to empower CNC. We start by analyzing the forces driving the development of CNC from the perspectives of influence of social forces and technical trends, followed by illustrations of a large number of new use cases. Then we discuss the connotation of CNC from the aspects of definition and convergence levels. Afterwards, the framework of distributed cooperative computing request routing is presented in detail, including trading plane, management plane, control plane and forwarding plane. The cross-plane cooperation routing schemes of the framework are analyzed to achieve deadline-aware end-to-end scheduling for the computing requests. Simulations are conducted to illustrate the efficiency of our routing schemes for ensuring performance of CNC systems.

\bibliographystyle{IEEEtranbst}
\bibliography{myref}

\begin{IEEEbiography}[{\includegraphics[width=1in,height=1.25in,clip,keepaspectratio]{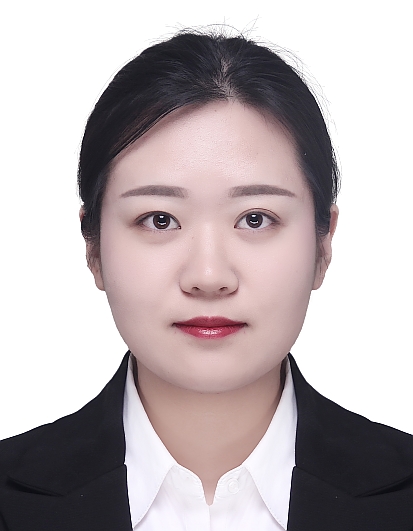}}]{Yujiao Hu}
    (Memeber, IEEE) received her Bachelor and PhD degrees from the Department of Computer Science of Northwestern Polytechnical University, Xi'an, China, in 2016 and 2021 respectively. From Nov. 2018 to March 2020, she was a visiting PhD student in National University of Singapore. Currently, she is a faculty member in Purple Mountain Laboratories. She focuses on computing and networks convergence, edge computing, deep learning, multi-agent cooperation, time sensitive networks and Internet of Things.     
\end{IEEEbiography}

\begin{IEEEbiography}[{\includegraphics[width=1in,height=1.25in,clip,keepaspectratio]{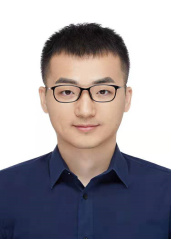}}]{Qingmin Jia}
	(Memeber, IEEE) is currently a Researcher in Future Network Research Center of Purple Mountain Laboratories. He received the B.S. degree from Qingdao University of Technology in 2014, and received the Ph.D. degree from Beijing University of Posts and Telecommunications (BUPT) in 2019. His current research interests include edge computing, edge Intelligence, software-defined network and Future Network Architecture. He has served as a Technical Program Committee Member of IEEE GLOBECOM 2021, HotICN 2021.
\end{IEEEbiography}

\begin{IEEEbiography}[{\includegraphics[width=1in,height=1.25in,clip,keepaspectratio]{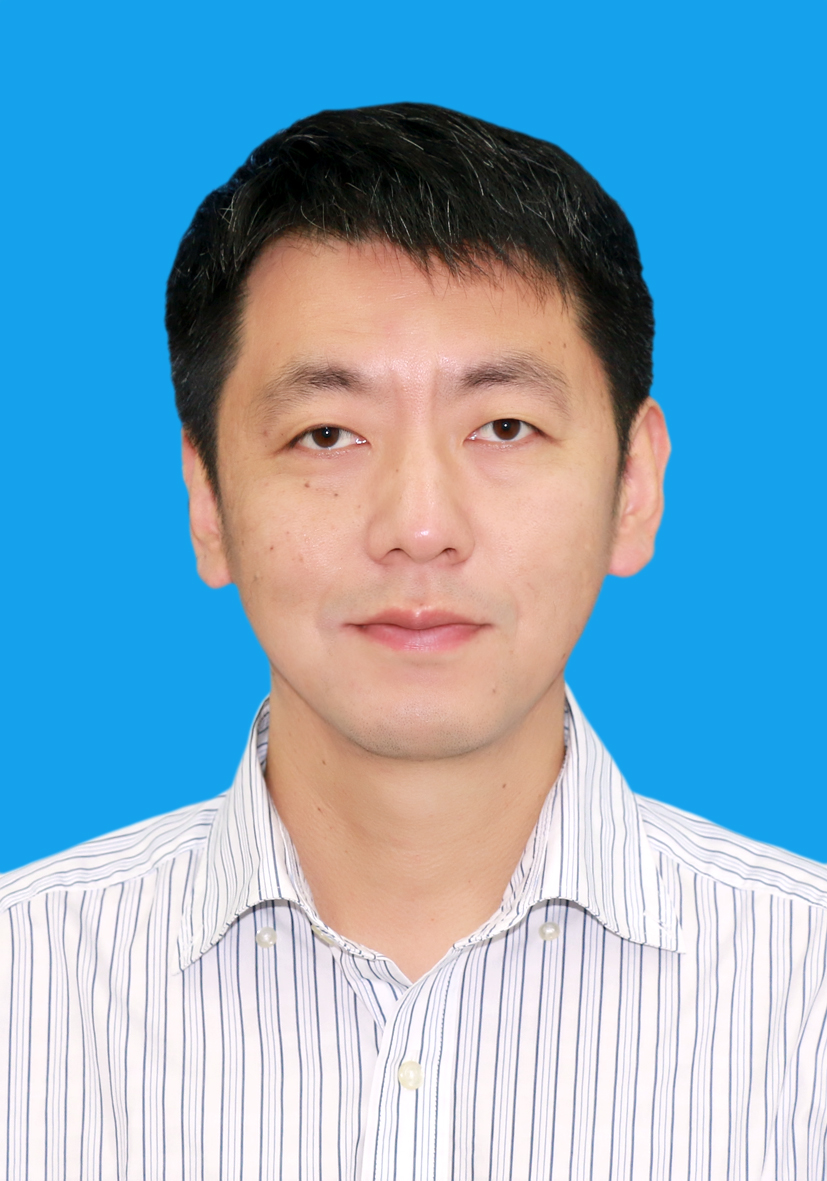}}]{Meng Shen} (Memeber, IEEE) received his Master and PhD degrees from the Department of Information and Communications Engineering of SouthEast University NanJing, China, in 2002 and 2021 respectively. From 2010 to 2021, he successfully founded several companies in the fields of communications IC and systems. Currently, he is a senior communication system expert in Purple Mountain Laboratories directing researches in the fields of B5G/6G technology evolution, low orbit satellite communications, AI driven multi-agent cooperation, time sensitive networks and Internet of Things. 
\end{IEEEbiography}

\begin{IEEEbiography}[{\includegraphics[width=1in,height=1.25in,clip,keepaspectratio]{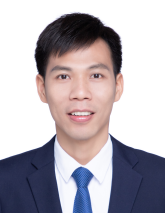}}]{Renchao Xie} 
	(Senior Member, IEEE) received the Ph.D. degree from the School of Information and Communication Engineering, Beijing University of Posts and Telecommunications (BUPT), Beijing, China, in 2012. He is a Professor with BUPT. From July 2012 to September 2014, he worked as a Postdoctoral Fellow with China United Network Communications Group Company. From November 2010 to November 2011, he visited Carleton University, Ottawa, ON, Canada, as a Visiting Scholar. His current research interests include 5G network and edge computing, information-centric networking, and future network architecture. Dr. Xie has served as a Technical Program Committee Member of numerous conferences, including IEEE Globecom, IEEE ICC, EAI Chinacom, and IEEE VTC-Spring.
\end{IEEEbiography}

\begin{IEEEbiography}[{\includegraphics[width=1in,height=1.25in,clip,keepaspectratio]{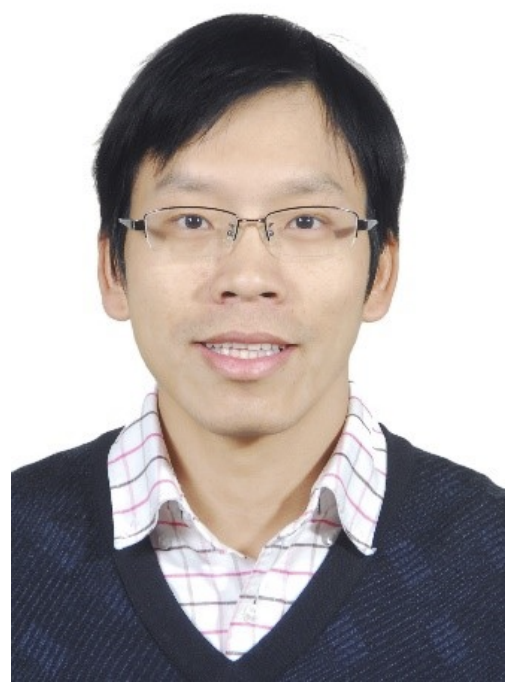}}]{Tao Huang} 
	(Senior Member, IEEE) received his B.S. degree in communication engineering from Nankai University, Tianjin, China, in 2002, the M.S. and Ph.D. degree in communication and information system from Beijing University of Posts and Telecommunications, Beijing, China, in 2004 and 2007 respectively. He is currently a professor at Beijing University of Posts and Telecommunications. His current research interests include network architecture, network artificial intelligence, routing and forwarding, and network virtualization.
\end{IEEEbiography}

\begin{IEEEbiography}[{\includegraphics[width=1in,height=1.25in,clip,keepaspectratio]{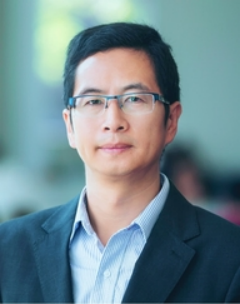}}]{F. Richard Yu } 
	(Fellow, IEEE) received the Ph.D. degree in electrical engineering from the University of British Columbia, Vancouver, BC, Canada, in 2003. From 2002 to 2006, he was with Ericsson, Lund, Sweden, and a start-up in California, USA. He joined Carleton University, Ottawa, ON, Canada, in 2007, where he is currently a Professor. His research interests include wireless cyber–physical systems, connected/autonomous vehicles, security, distributed ledger technology, and deep learning. He is a Distinguished Lecturer, the Vice President (Membership), and an Elected Member of the Board of Governors of the IEEE Vehicular Technology Society. He is a Fellow of the IEEE, Canadian Academy of Engineering (CAE), Engineering Institute of Canada (EIC), and Institution of Engineering and Technology (IET).
\end{IEEEbiography}

\end{document}